# Social Network Sensors for Early Detection of Contagious Outbreaks

Nicholas A. Christakis[1*], James H. Fowler[2]

[1]Harvard Faculty of Arts and Sciences and Harvard Medical School, Boston, MA 02115, USA
[2]Political Science Department, University of California, San Diego, La Jolla, CA 92103, USA
* To whom correspondence should be addressed, email: Nicholas_christakis@harvard.edu, phone: (617) 432-5890.

**Current methods for the detection of contagious outbreaks give contemporaneous information about the course of an epidemic at best. Individuals at the center of a social network are likely to be infected sooner, on average, than those at the periphery. However, mapping a whole network to identify central individuals whom to monitor is typically very difficult. We propose an alternative strategy that does not require ascertainment of global network structure, namely, monitoring the friends of randomly selected individuals. Such individuals are known to be more central. To evaluate whether such a friend group could indeed provide early detection, we studied a flu outbreak at Harvard College in late 2009. We followed 744 students divided between a random group and a friend group. Based on clinical diagnoses, the progression of the epidemic in the friend group occurred 14.7 days (95% C.I. 11.7–17.6) in advance of the randomly chosen group (i.e., the population as a whole). The friend group also showed a significant lead time ($p<0.05$) on day 16 of the epidemic, a full 46 days before the peak in daily incidence in the population as a whole. This sensor method could provide significant additional time to react to epidemics in small or large populations under surveillance. Moreover, the method could in principle be generalized to other biological, psychological, informational, or behavioral contagions that spread in networks.**

**Introduction**

Current methods for the detection of contagious outbreaks ideally give contemporaneous information about the course of an epidemic, though, more typically, they lag behind the epidemic.(*1-3*) However, the situation could be improved, possibly significantly, if detection methods took advantage of a potentially informative property of social networks: during a contagious outbreak, individuals at the center of a network are likely to be infected sooner than random members of the population. Hence, the careful collection of information from human social networks could be used to detect contagious outbreaks *before* they happen in the general population.

A contagion that stochastically infects some individuals and then spreads from person to person in the network will tend, on average, to reach centrally-located individuals more quickly than peripheral individuals because central individuals are a smaller number of steps (degrees of separation) away from the average individual in the network (see Figure 1).(*4,5*) Although some contagions can spread via incidental contact, the duration of exposure between people with social ties is typically much higher than between strangers, suggesting that the social network will be an important conduit for the course of an outbreak.(*5,6*) As a result, we would expect the S-shaped epidemic curve (*7,8*) to be shifted to the left (forward in time) for centrally located individuals compared to the population as a whole (see Figure 2). This shift, if it could be observed, would allow for early detection of an outbreak.

Prior modeling research suggests that vaccinating central individuals in networks could enhance the population-level efficacy of a prophylactic intervention (*9-12*) and that optimal placement of sensors in physical networks (such as water pumping stations) could detect outbreaks sooner.(*13*) However, mapping a whole network to identify particular individuals from whom to collect information is costly, time-consuming, and often impossible, especially for large networks. We therefore explore a novel, alternative strategy that does *not* require ascertainment of global network structure, namely, *surveying the friends of randomly selected* *individuals*. This strategy exploits an interesting property of human social networks: on average, the friends of randomly selected people possess more links (have higher degree) and are also more central to the network than the initial, randomly selected people who named them.(*14-18*) Therefore, we expect a set of nominated friends to get infected earlier than randomly chosen individuals (who represent the population as a whole).

To our knowledge, a method that uses nominated friends as sensors for early detection of an outbreak has not previously been proposed or tested on any sort of real outbreak. To evaluate the effectiveness of nominated friends as social network sensors, we therefore monitored the spread of flu at Harvard College from September 1 to December 31, 2009. In the fall of 2009, both seasonal flu (which typically kills 41,000 Americans each year (*19*)) and the H1N1 strain were prevalent in the US, though the great majority of cases in 2009 have been attributed to the latter.(*1*) It is estimated that the H1N1 epidemic, which began roughly in April 2009, has infected over 50 million Americans. Unlike seasonal flu, which typically affects individuals older than 65, H1N1 tends to affect young people. Nationally, according to the CDC, the epidemic peaked in late October 2009, and vaccination only became widely available in December 2009. Whether another outbreak of H1N1 will occur (for example, in areas and populations that have heretofore been spared) is a matter of some debate at present,(*1*) but many scholars have been studying the situation from biological and public health perspectives.(*20,21*)

We enrolled a total of 744 undergraduate students from Harvard College, discerned their friendship ties, and tracked whether they had the flu beginning on September 1, 2009 (from the start of the new academic year) to December 31, 2009. This sample was assembled by empanelling two groups of students of essential analytic interest here: a "random" sample chosen randomly from the 6,650 Harvard undergraduates (N=319) and a "friends" sample (N=425) composed of individuals who were named as a friend at least once by a member of the random group.

In addition, as a byproduct of empanelling the foregoing group of 744 students, we acquired information about a total of 1,789 uniquely identified students (who either participated in the study or who were nominated as friends or as friends of friends) with which to draw the social network of part of the Harvard College student body. Our sample of 744 was thus embedded in this larger network of 1,789 people (see SI for more details).

After giving informed consent, all subjects completed a brief background questionnaire soliciting demographic information, flu and vaccination status since September 1, 2009, and certain self-reported measures of popularity. We also obtained basic administrative data from the Harvard College registrar, such as sex, class of enrollment, and sports participation.

We tracked cases of formally diagnosed influenza among the students in our sample as recorded by University Health Services (UHS) beginning on September 1, 2009 through December 31, 2009. Presenting to the health service indicates a more severe level of symptomatology, of course, and so we do not expect the same overall prevalence using this diagnostic standard as with self-reported flu discussed below. However, UHS data offer the advantage of allowing us to obtain information about flu symptoms as assessed by medical staff.

Beginning on October 23, 2009, we also collected self-reported flu symptom information from participants via email twice weekly (on Mondays and Thursdays), continuing until December 31, 2009. The students were queried about whether they had had a fever or flu symptoms since the last email contact, and there was very little missing data (47% of the subjects completed *all* of the biweekly surveys, and 90% missed no more than two of the surveys).

Self-report of symptoms rather than serological testing is the current standard for flu diagnosis. Similar to previous studies,(*22*) students were deemed to have a case of flu (whether seasonal or the H1N1 variety) if they reported having a fever of greater than 100° F (37.8° C) *and* at least two of the following symptoms: sore throat; cough; stuffy or runny nose; body aches; headache; chills; or fatigue. We checked the sensitivity of our findings by using definitions of flu that required more symptoms, and our results did not change (see SI). As part of the foregoing biweekly self-reports, in order to complement the UHS vaccination records, we also ascertained whether the students reported having been vaccinated (with seasonal flu vaccine or H1N1 vaccine or both) at places other than (and including) UHS.

**Results**

By December 31, 2009, the cumulative incidence of flu in our sample was 8% based on diagnoses by medical staff, and it was 32% based on self-reports, which mirrored other studies of school-based outbreaks and also recent national estimates for the college-student population.(*22,23*)  As expected, the prevalence was higher by the latter standard.  We studied the association of several demographic and other variables with cumulative flu incidence at day 122 (the last day of follow-up) to see whether they predicted an increase in overall risk.  None of these variables was significantly associated with flu diagnoses by medical staff (see SI), so we focused on the effect of these variables on shifts in the timing of the distribution.

As hypothesized, the cumulative incidence curves for the friend group and the random group diverge and then converge (Figure 3).  NLS estimates suggest that the friends curve for flu diagnosed by medical staff is shifted 14.7 days forward in time (95% C.I. 11.7–17.6), thus providing early detection.  This represents approximately 65% of one standard deviation in the time to event in the whole sample.  The results also indicate a significant but smaller shift in self-reported flu symptoms (3.2 days, 95% C.I. 2.2–4.3).  In both the clinical and self-reported diagnostic standards, the estimates are robust to a number of control variables including H1N1 vaccination, seasonal flu vaccination, sex, college class, and varsity sports participation (see SI).

The foregoing estimates rely on full information *ex post*, but we wondered when it would also be possible to detect a difference in the friend group and the random group in real time, given less complete data.  We therefore estimated the models each day using all available information up to that day.  For flu diagnoses by medical staff, the friend group showed a significant lead time ($p<0.05$) on day 16, a full 46 days before the estimated peak in daily incidence in visits to the health service. For self-reported flu symptoms, the friend group showed a significant lead time by day 39, which is 83 days prior to the estimated peak in daily incidence in self-reported symptoms. Thus, a comparison of outcomes in friends and randomly chosen individuals could be an additional effective technique for detecting outbreaks at early stages of an epidemic.

A possible alternative to the friendship nomination procedure would be to rely on self-reported popularity or self-reported counts of numbers of friends in order to identify a high-risk group. We measured our subjects' self-perceptions of popularity using an eight-item scale, but this did not yield a significant shift forward in time for flu diagnoses (see SI). Moreover, controlling for self-reported popularity did not alter the significance of the lead time provided by the friend group for either flu diagnoses by medical staff or self-reported flu symptoms. These results suggest that being nominated as a friend captures more network information (including the tendency to be central in the network) than self-reported network attributes.

Although the method described here does not require information about the full network, our survey took place on a college campus in which many nominators were themselves nominated, and the same person was frequently nominated several times. As a result, a connected component of 714 people emerged out of the 1,789 unique individuals who were either surveyed or identified as friends by those who took part in the study. We illustrate the spread of flu in this component in Figure 4, which shows the tendency of the flu to "bloom" in more central nodes of the network, and also in a 122-frame movie of daily flu prevalence available online (see SI).

Sampling a densely interconnected population also allowed us to measure egocentric network properties like in-degree (number of times a subject was nominated as a friend), betweenness centrality (the number of shortest paths in the network that pass through an individual), and transitivity (the probability that two of one's friends are friends with one another). The results showed that, as expected, the friend group differed significantly from the random group for all these measures, exhibiting higher in-degree (Mann Whitney U test $p<0.001$), higher centrality ($p<0.001$), and lower transitivity ($p=0.039$).

We hypothesized that each of these measures could help to identify groups that could be used as social network sensors when full network information is, indeed, available (see Figure 5). For example, we expect in-degree to be associated with early contagion because more friends mean more paths to others in the network who might be infected. NLS estimates suggest that each additional nomination shifts the flu curve left by 5.6 days (95% C.I. 3.6–8.1) for flu diagnoses by medical staff and 8.0 days (95% C.I. 7.3–8.5) for self-reported symptoms. On the other hand, the same is not true for out-degree (the number of friends a person names); pertinently, this is the only quantity that would be straightforwardly ascertainable by asking respondents about themselves. However, there is low variance in this measure in the present setting since most people named three friends.

We also expect betweenness centrality to be associated with early contagion. NLS estimates suggest that individuals with maximum observed centrality shift the flu curve left by 16.5 days (95% C.I. 1.9–28.3) for flu diagnoses by medical staff and 22.9 days (95% C.I. 20.0–27.2) for self-reported symptoms, relative to those with minimum centrality. Moreover, centrality remains significant even when controlling for both in-degree and out-degree, suggesting that it is not just the number of friends that is important, but also the number of friends of friends, friends of friends of friends, and so on.

Finally, we expect transitivity to be negatively associated with early contagion. People with high transitivity may be poorly connected to the rest of the network because their friends tend to know one another and exist in a tightly-knit group. In contrast, those with low transitivity tend to be connected to many different, independent groups, and each additional group increases the possibility that someone in that group has the flu and that it spreads to the subject. NLS estimates suggest that individuals with minimum observed transitivity shift the flu curve left by 31.9 days (95% C.I. 23.5–43.5) for flu diagnoses by medical staff and 15.0 days (95% C.I. 12.7–18.5) for self-reported symptoms compared to those with maximum transitivity. Moreover, transitivity remains significant even when controlling for both in-degree and out-degree.

**Discussion**

For many contagious diseases, early knowledge of when – or whether – an epidemic is unfolding is crucial to policy makers and public health officials responsible for defined populations, whether small or large. In fact, with respect to flu, models assessing the impact of prophylactic vaccination in a metropolis such as New York City suggest that vaccinating even one third of the population would save lives and shorten the course of the epidemic, but only if implemented a month earlier than usual.(*24,25*) A method like the one described here could help provide such early warning.

In fact, this method could be used to monitor targeted populations of any size, in real time. For example, a health service at a university (or other institution) could empanel a sample of subjects who are nominated as friends and who agree to be passively monitored for their health care use; a spike in cases in this group could be read as a warning of an impending outbreak. Public health officials responsible for a city could empanel a sample of randomly chosen individuals and a sample of nominated friends (perhaps a thousand people in all) who have agreed to report their symptoms using brief, periodic text messages or an online survey system (like the one employed here). Regional or national populations could also be monitored in this fashion, with a sample of nominated friends being periodically surveyed instead of, or in addition to, a random sample of people (as is usually the norm). Since public health officials often monitor populations in any case, the change in practice required to monitor a sample of these more central individuals would not be too burdensome.

Moreover, whereas officials responsible for a single, relatively small institution might possibly actively seek out central individuals to vaccinate them (hence potentially confounding the utility of such individuals as sensors), such a vaccination effort would be unlikely to be initiated with a regional or national sample, given the likely irrelevance of vaccinating the actual sensor sample members as a means to control any wide-scale epidemic.

The difference in the timing of the course of the epidemic in the friend and random groups could be exploited in at least two different ways. First, if solely the friends group were being monitored, an analyst tracking an outbreak might look for the first evidence that the incidence of the pathogen among the friends group rose above a predetermined rate (e.g., a noticeable increase above a zero background rate); this itself could indicate an impending epidemic. Second, in a strategy that would yield more information, the analyst could track both a sample of friends and a sample of random subjects, and the harbinger of an epidemic could be taken to be when the two curves were seen to first diverge from each other. Especially in the case of the spread of contagions other than biological pathogens, the difference between these two curves provides additional information: the adoption curve among the random sample provides evidence of secular trends in the population, whereas the *difference* between the two curves provides evidence of a network effect, over and above the baseline force of the epidemic.

While our goal here was to evaluate how the method of surveying friends could provide early detection of contagious outbreaks in general, it is noteworthy that, in the specific case of the flu, the method we evaluated appears to provide longer lead times than other extant methods of monitoring flu epidemics. Current surveillance methods for the flu, such as those implemented by the CDC that require collection of data from subjects seeking outpatient care or having lab tests, are typically lagging indicators about the timing of the epidemic (information is typically one to two weeks behind the actual course).(*1*) A proposal to use Google Trends to monitor searches for information about flu suggests that this approach could offer a better indicator, providing evidence of an outbreak at least a week before published CDC reports.(*2,3*) However, while potentially instantaneous, the Google Trends method would only, at best, give *contemporaneous* information about rates of infection (plus, the search algorithm would have to be customized for each pathogen of interest). In contrast, we show that the sensor method described here can detect an outbreak of flu two weeks *in advance*. That is, the sensor network method provides early detection rather than just rapid warning.

Moreover, the sensor method could be used in conjunction with online search. By following the online behavior of a friend group, or a group known to be central in a network (for example, based on e-mail records which could be used to reconstruct social networks), Google or other search engines might be able to get high-quality, real-time information about the epidemic with even greater lead time, giving public health officials even more time to plan a response.

How much advance detection would be achieved for other pathogens or in populations of larger size or different composition remains unknown. The ability of the proposed method to detect outbreaks early, and how early it might do so, will depend on intrinsic properties of the thing that is spreading (e.g., the biology of the pathogen); how that thing is measured; the nature of the population, including the overall prevalence of susceptible or affected individuals; the number of people empanelled into the sensor group; the topology of the network (for example, the degree distribution and its variance, or other structural attributes;(*26*) and other factors, such as whether the outbreak modifies the structure of the network as it spreads (for example, by killing people in the network, or, in the case of spreading information, perhaps by affecting the tendency of any two individuals to remain connected after the information is transmitted).

While the social network sensor strategy has been illustrated with a particular outbreak (flu) in a particular population (college students), it could potentially be generalized to other phenomena that spread in networks, whether biological (antibiotic-resistant germs), psychological (depression), normative (altruism),(*27*) informational (rumors), or behavioral (smoking).(*28*) Outbreaks of a wide variety of deleterious or desirable conditions could be detected before they have reached a critical threshold in populations of interest.

**Materials and Methods**

To measure self-perceived popularity, we adapted a set of 8 questions previously used to assess the popularity of co-workers.(*29*)

We used friendship nominations to measure the *in-degree* (the number of times an individual is named as a friend by other individuals) and *out-degree* (the number of individuals each person names as a friend) of each subject. The in-degree is virtually unrestricted (the theoretical maximum is $N-1$, the total number of other people in the network) but the out-degree is restricted to a maximum of 3, given the way we elicited friendship information.

We measured *betweenness centrality*, which identifies the extent to which an individual lies on potential paths for contagions passing from one individual to another through the network; this quantity summarizes how central an individual is in the network (see Figure 1).(*30*) We measured *transitivity* as the empirical probability that two of a subject's friends are also friends with each other, forming a triangle (see Figure 1). This measure is just the total number of triangles of ties between an individual and his or her social contacts divided by the total possible number of triangles.

We used Pajek (*31*) to draw two-dimensional pictures of the network, and we implemented the Kamada-Kawai algorithm, which generates a matrix of shortest network path distances from each node to all other nodes in the network and repositions nodes in an image so as to reduce the sum of the difference between the plotted distances and the network distances.(*32*) A movie of the spread of flu with a frame for each of the 122 days of the study is available online (see SI).

We calculated the cumulative flu incidence for both the friend group and the random group using a nonparametric maximum likelihood estimate (NPMLE).(*33*) We also calculated the predicted daily incidence using an estimation procedure designed to measure the shift in the time course of a contagious outbreak associated with a given independent variable (see SI). In this procedure, we fit the observed probability of flu to a cumulative logistic function via nonlinear least squares (NLS) estimation.(*34*) To derive standard errors and 95% confidence intervals, we used a bootstrapping procedure in which we repeatedly re-sampled subject observations with replacement and re-estimated the fit. (*35*)  This procedure produced somewhat wider confidence intervals than those based on asymptotic approximations, so we report only the more conservative bootstrapped estimates.  Finally, we calculated how many days of early detection was possible for groups with various network attributes by multiplying the coefficient and confidence intervals in the foregoing models by the mean difference between the above-average group and the below-average group (see SI).

**Acknowledgements**

Supported by a grant from the Office of the Provost, Harvard University (Steve Hyman). The analysis was partially supported by grants from the National Institute on Aging (P-01 AG-031093) and by a Pioneer Grant from the Robert Wood Johnson Foundation.  We thank David Rosenthal and Paul Barreira for advice about flu at Harvard College and for access to the University Health Services data.  We thank Paul Allison, Guido Imbens, and Jukka-Pekka Onnela for helpful comments on the manuscript.  We thank Weihua An, Tom Keegan, Mark McKnight, Laurie Meneades, and Alison Wheeler for help with data collection and preparation.  And we thank all the students who contributed their time and consented to having their health records examined.

22. J. Lessler, N.G. Reich, and D.A.T. Cummings. Outbreak of 2009 Pandemic Influenza A (H1N1) at a New York City School. *New Engl J Med* 2009; 361: 2628–2636.

23. ACHA. *American College Health Association Influenza Like Illnesses (ILI) Surveillance in Colleges and Universities 2009-2010: Weekly College ILI cases reported* (Linthicum, MD: American College Health Association; 2010).

24. N. Khazen, D.W. Hutton, A.M. Garber, N. Hupert, DK. Owens. Effectiveness and cost-effectiveness of vaccination against pandemic influenza (H1N1). *Ann Intern Med* 2009; 151: 829–839.

25. V.J. Davey, R.J. Glass, H.J. Min, W.E. Beyeler, L.M. Glass. Effective, robust design of community mitigation for pandemic influenza: A systematic examination of proposed US guidance. *PLoS ONE* 2008; 3(7): e2606. doi:10.1371/journal.pone.0002606.

26. P.S. Bearman, J. Moody, and K. Stovel. Chains of affection: The structure of adolescent romantic and sexual networks. *Am J Sociol* 2004; 110: 44-91

27. J.H. Fowler and N.A. Christakis. Cooperative Behavior Cascades in Human Social Networks. *PNAS* 2010; 107: 5334-5338.

28. N.A. Christakis and J.H. Fowler. The collective dynamics of smoking in a large social network. *New Engl J Med* 2008; 358: 2249–2258.

29. B.A. Scott, T.A. Judge. The popularity contest at work: who wins, why, and what do they receive? *J Applied Psych* 2009; 94: 20–33.

30. L.C. Freeman. Set of measures of centrality based on betweenness. *Sociometry* 1977; 40:35–41.

31. V. Batagelj, A. Mrvar. Program for Analysis and Visualization of Large Networks, version 1.14, 2006.
15

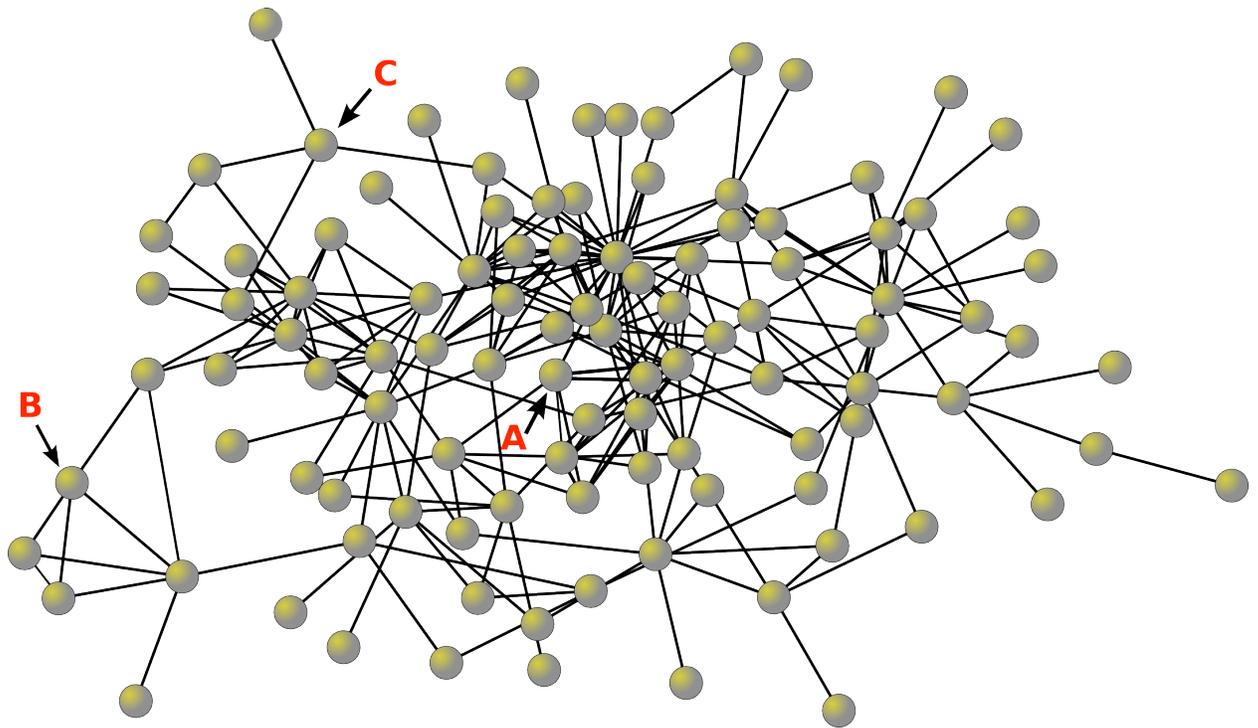

**Figure 1. Network Illustrating Structural Parameters.** This real network of 105 students shows variation in structural attributes and topological position. Each circle represents a person and each line represents a friendship tie. Nodes A and B have different "degree," a measure that indicates the number of ties. Nodes with higher degree also tend to exhibit higher "centrality" (node A with six friends is more central than B and C who both only have four friends). If contagions infect people at random at the beginning of an epidemic, central individuals are likely to be infected sooner because they lie a shorter number of steps (on average) from all other individuals in the network. Finally, although nodes B and C have the same degree, they differ in "transitivity" (the probability that any two of one's friends are friends with each other). Node B exhibits high transitivity with many friends that know one another. In contrast, node C's friends are not connected to one another and therefore they offer more independent possibilities for becoming infected earlier in the epidemic.

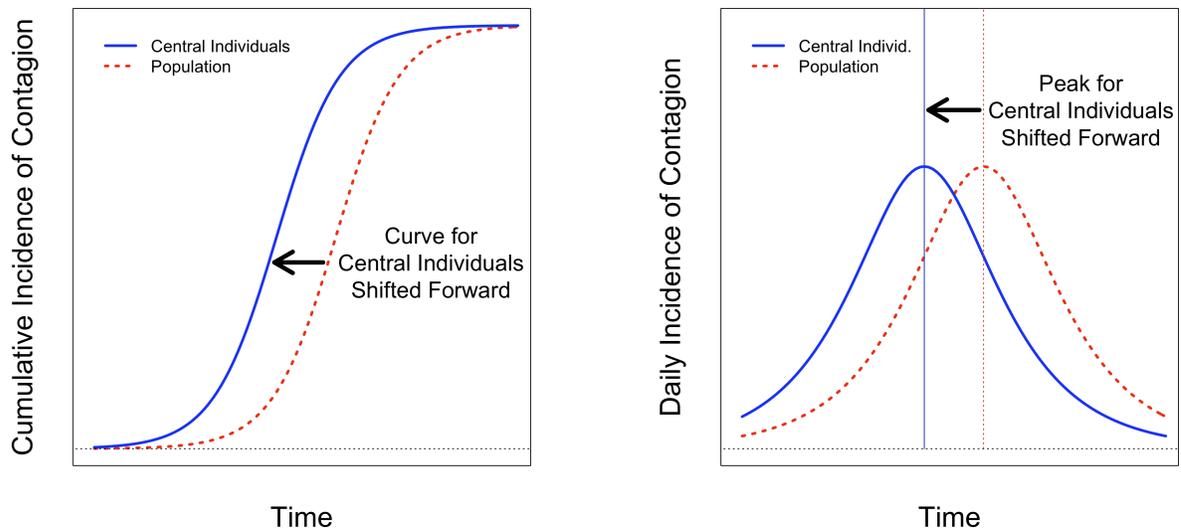

**Figure 2. Theoretical expectations of differences in contagion between central individuals and the population as a whole.** A contagious process passes through two phases, one in which the number of infected individuals exponentially increases as the contagion spreads, and one in which incidence exponentially decreases as susceptible individuals become increasingly scarce. These dynamics can be modelled by a logistic function. Central individuals lie on more paths in a network compared to the average person in a population and are therefore more likely to be infected early by a contagion that randomly infects some individuals and then spreads from person to person within the network. This shifts the S-shaped logistic cumulative incidence function forward in time for central individuals compared to the population as a whole (left panel). It also shifts the peak infection rate forward (right panel).

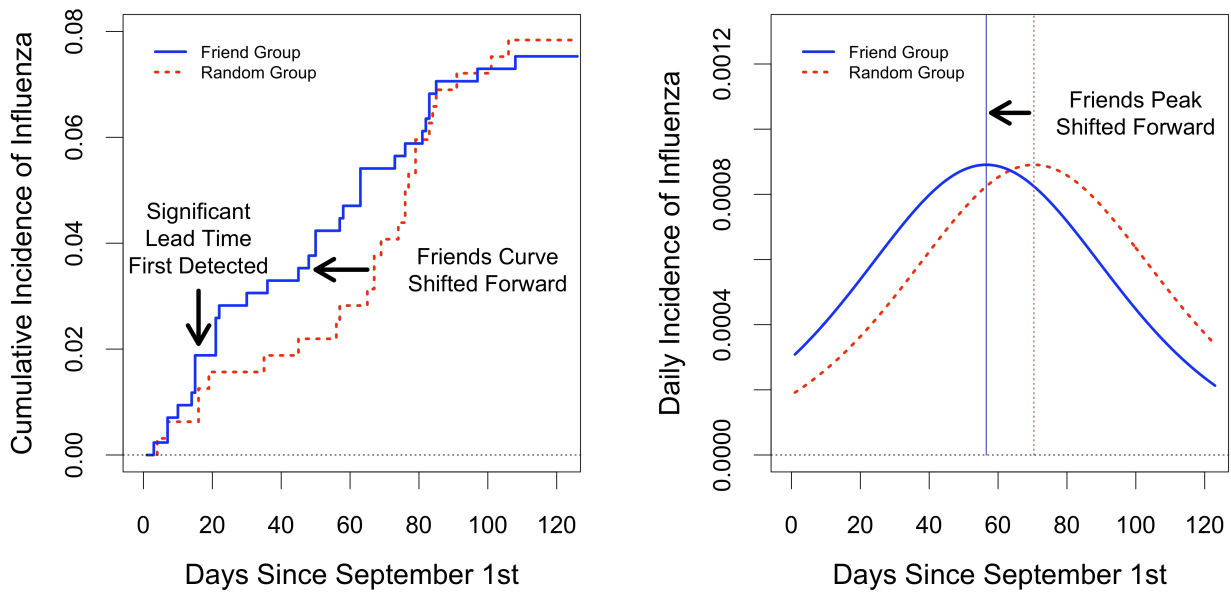

**Figure 3. Empirical differences in flu contagion between "friend" group and randomly chosen individuals.** We compared two groups, one composed of individuals randomly selected from our population, and one composed of individuals who were nominated as a friend by members of the random group. The friend group was observed to have significantly higher measured in-degree and betweenness centrality than the random group (see SI). In the left panel, a nonparametric maximum likelihood estimate (NPMLE) of cumulative flu incidence (based on diagnoses by medical staff) shows that individuals in the friend group tended to get the flu earlier than individuals in the random group. Moreover, predicted daily incidence from a nonlinear least squares fit of the data to a logistic distribution function suggests that the peak incidence of flu is shifted forward in time for the friends group by 14.7 days (right panel). A significant ($p<0.05$) lead time for the friend group was first detected with data available up to Day 16.

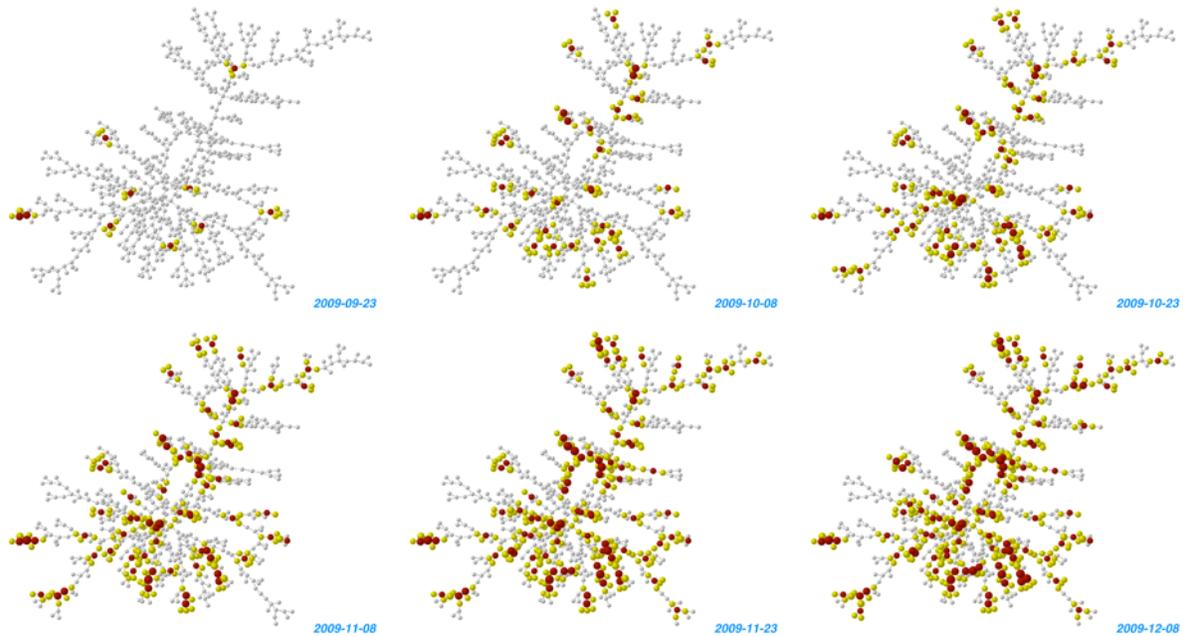

**Figure 4. Progression of flu contagion in the friendship network over time.** Each frame shows the largest component of the network (714 people) for a specific date, with each line representing a friendship nomination and each node representing a person. Infected individuals are colored red, friends of infected individuals are colored yellow, and node size is proportional to the number of friends infected. All available information regarding infections is used here. Frames for all 122 days of the study are available in a movie of the epidemic posted in the Supplementary Information.

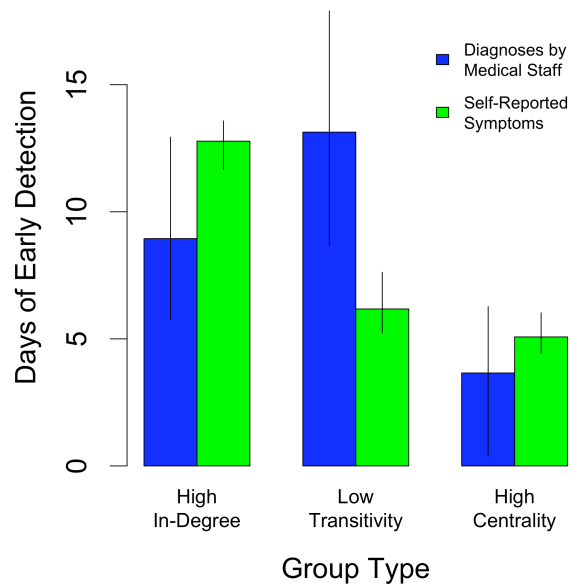

**Figure 5. Estimated days of advance detection of a flu outbreak when following specific groups.** Here, degree, transitivity, and centrality are computed based on the mapping of the network. The high in-degree group is composed of individuals who have a higher-than-average number of other people in the network who name them as a friend. The low transitivity group is composed of individuals with below-average probability that any two of their friends are friends with one another. The high centrality group is composed of individuals with a higher-than-average betweenness, which is the number of shortest paths connecting all individuals in a network that pass through a given person. Analyses were conducted separately for data based on flu diagnoses by medical staff (blue bars) and data based on self-reported flu symptoms (green bars). Estimates and 95% confidence intervals are based on a nonlinear least squares fit of the flu data to a logistic distribution function (see SI). The results show that flu outbreaks occur up to two weeks earlier in each of these groups.

# Social Network Sensors for Early Detection of Contagious Outbreaks

Nicholas A. Christakis[1], James H. Fowler [2]

[1] *Harvard Medical School and Harvard Faculty of Arts and Sciences, Boston, MA 02115, USA*

[2] *Political Science Department, University of California, San Diego, La Jolla, CA 92103, USA*

Correspondence and requests for materials should be addressed to N.A.C. (e-mail: christakis@hcp.med.harvard.edu, phone: 617-432-5890.

**Subjects**

We enrolled a total of 744 undergraduate students from Harvard College, discerned their friendship ties, and tracked whether they had the flu beginning on September 1, 2009, from the start of the new academic year, to December 31, 2009.

Beginning on October 23, 2009, we approached 1,300 randomly selected Harvard College students (out of 6,650); we waited until a few weeks of the new school year had passed in order to be able to obtain current friendship information. Of these 1,300 students, 396 (30%) agreed to participate. All of these students were in turn asked to nominate up to three friends, and a total of 1,018 friends were nominated (average of 2.6 friends per nominator). This yielded 950 unique individuals to whom we sent the same invitation as the initial group. Of these, 425 (45%) agreed to participate. However, 77 of these 950 subjects were themselves members of the original, randomly selected group and hence were already participants. Thus, the sample size after the enrolment of the random group and the friend group was 744.

Nominated friends were sent the same survey as their nominators; hence, the original 425 friends also nominated 1,180 of their own friends (average of 2.8 friends per nominator), yielding 1004 further, unique individuals. Although we did *not* send surveys to these "friends of friends," 303 (30%) were themselves already enrolled either in the friends group or in the initial randomly selected group.

Thus, in the end, we have empanelled two groups of students of essential analytic interest here: a "random" sample (N=319) and a "friends" sample (N=425) composed of individuals who were named as a friend at least once by a member of the random group. In addition, we ultimately had information about a total of 1,789 uniquely identified students (who either participated in the study or who were nominated as friends or friends of friends) with which to draw social networks of the Harvard College student body (27% of all 6,650 undergraduates). Our sample of 744 was thus embedded in this larger network of 1,789 people.

After giving informed consent, all subjects completed a brief background questionnaire soliciting demographic information, flu and vaccination status since September 1, 2009, and certain self-reported measures of popularity. We also obtained basic administrative data from the Harvard College registrar, such as sex, class of enrolment, and information about participation in varsity sports.

We also tracked cases of formally diagnosed influenza among the students in our sample as recorded by University Health Services (UHS) beginning on September 1, 2009 through December 31, 2009. Presenting to the health service indicates a more severe level of symptomatology, of course, and so we do not expect the same overall prevalence using this diagnostic standard as with self-reported flu discussed below. However, UHS data offer the advantage of allowing us to obtain information about flu symptoms as assessed by medical staff. A total of 627 of the 744 students (84%) who agreed to participate in the survey portion of our study also gave written permission for us to obtain their health records. Finally, 7 students reported being diagnosed with flu by medical staff at facilities other than UHS (in response to survey questions asked of all students), so we include these in the data as well.

Notably, we do not expect cases of flu to meaningfully alter the social networks and friendship patterns of Harvard undergraduates, let alone over a two-month period. And, we assume that the friendship network of Harvard students in our sample did not change meaningfully over the period September to December. That is, we treat the network as static over this time interval.

Beginning on October 23, 2009, we also collected *self-reported* flu symptom information from participants via email twice weekly (on Mondays and Thursdays), continuing until December 31, 2009. The enrolled students were queried about whether they had had a fever or flu symptoms since our last email contact, and there was very little missing data (47% of the subjects completed *all* of the biweekly surveys, and 90% missed no more than two of the surveys).

Self-report of symptoms rather than serological testing is the current standard for flu diagnosis. Students were deemed to have a case of flu (whether seasonal or the H1N1 variety) if they report having a fever of greater than 100° F (37.8° C) *and* at least two of the following symptoms: sore throat; cough; stuffy or runny nose; body aches; headache; chills; or fatigue. We checked the sensitivity of our findings by using definitions of flu that required more symptoms, and our results did not change. As part of the foregoing biweekly follow-up, and to supplement the UHS vaccination records, we also ascertained whether the students reported having been vaccinated (with seasonal flu vaccine or H1N1 vaccine or both) at places other than (and including) UHS.

Hence, we had two measures of flu incidence. The medical-staff standard was a formal diagnosis by a health professional and typically reflected more severe symptoms. The self-reported standard captured cases that did not come to formal medical attention. As expected, the cumulative incidence of the latter was approximately four times the former (32% versus 8%) by the time of cessation of follow-up on December 31, 2009.

**Network Measures**

We use friendship nominations to measure the *in-degree* (the number of times an individual is named as a friend by other individuals) and *out-degree* (the number of individuals each person names as a friend) of each subject. The in-degree is virtually unrestricted (the theoretical maximum is $N-1$, the total number of other people in the network) but the out-degree is restricted to a maximum of 3 due to the name generator used.

We also measure *transitivity* as the empirical probability that two of a subject's friends are also friends with each other, forming a triangle. This measure is just the total number of triangles of ties divided by the total possible number of triangles for each individual. This measure is undefined for individuals with less than 2 friends (23 cases out of 744), and so we treat this measure as missing in those cases.

Finally, we measure *betweenness centrality*, which identifies the extent to which an individual lies on potential paths for passing contagions from one individual to another through the network.[1] If we let $\sigma_{ik}$ represent the number of shortest paths from subject *i* to subject *k*, and $\sigma_{ijk}$ represent the number of shortest paths from subject $i$ to subject $k$ that pass through subject $j$, then the betweenness centrality measure $x$ for subject $j$ is $x_j = \sum_{i \neq j \neq k} \frac{\sigma_{ijk}}{\sigma_{ik}}$. To ease interpretability we divided all scores by $\max(x_j)$ so that all measures would lie between and including 0 and 1.

Note that for the purpose of measuring transitivity and betweenness centrality, we assume all directed ties are undirected, so that a tie in either direction becomes a mutual tie. For example, we consider the case where A names B, B names C, and C names A to be transitive. Likewise, if A names B, A names C, and B names C, we consider the relationships to be transitive for all three individuals.

We used Pajek[2] to draw pictures of the networks and used the Kamada-Kawai algorithm, which generates a matrix of shortest network path distances from each node to all other nodes in the network and repositions nodes so as to reduce the sum of the difference between the plotted distances and the network distances.[3] A movie of the spread of flu with a frame for each of the 122 days of the study is available online at http://jhfowler.ucsd.edu/flunet_v3.mov.

While it is the case that, in situations of *chronic* illness, people that are sick may have fewer friends or different network architectures as a result of their illness, we do not anticipate a problem with this phenomenon in this setting. That is, we do not think that undergraduate friendships will be modified by virtue of having the flu, especially over the short time intervals being studied here.

**Personality Measures**

To measure self-perceived popularity, we adapted a set of 8 questions previously used to assess the popularity of co-workers.[4] Specifically, we asked subjects to rate on a 5 point scale their agreement (ranging from *strongly disagree* to *strongly agree*) with the following statements: "I am popular," "I am quite accepted," "I am well-known," "I am generally admired," "I am liked," "I am socially visible," "I am viewed fondly," and "I am not popular" (reverse scored). We generated index scores via a one-dimension factor analysis of all 8 items (Cronbach's alpha=0.66).

**Analysis**

In Table S1 we report summary statistics for the random group and the friend group and the results of a Mann Whitney U test, which is a nonparametric test of differences in the two distributions. Notice that the friend group exhibits significantly higher in-degree and betweenness centrality, and significantly lower transitivity than the random group, as theorized. In addition, we find that the friend group has significantly more females and fewer sophomores than the random group.

In Table S2 we present Spearman correlations with *p* values to evaluate whether or not any study variables influence overall risk of getting the flu by December 31, 2009. Notice that the self-reported and medical staff measures are highly correlated at $\rho = 0.40$. However, no other variable is significantly associated with both measures. The two strongest associations with self-reported flu are in-degree and being a sophomore, but at 0.08 neither of these associations is strong and neither is confirmed in the data based on diagnoses by medical staff.

In Tables S3-S12, we report results from an estimation procedure designed to measure the shift in the time course of a contagious outbreak associated with a given independent variable. We fit the observed probability of flu to a cumulative logistic function

$$P_{it} = \lambda \left(1 + e^{\frac{-(t+\alpha+\mathbf{b}X_{it})}{\sigma}}\right)^{-1}$$

where $P_{it}$ is the probability subject $i$ has the flu on or before day $t$; $t + \alpha + \mathbf{b}X_{it}$ is a function that determines the location of peak risk to subject $i$ on day $t$ that includes a constant $\alpha$, a vector of coefficients $\mathbf{b}$, and a matrix of independent variables $X_{it}$; $\sigma$ is a constant scale factor that provides an estimate of the standard deviation in days of the time course of the epidemic; and $0 \leq \lambda \leq 1$ is a constant indicating the maximum cumulative risk. For medical diagnoses by staff, we assume $P_{it}$ is 1 when subjects have had the flu on any day up to and including $t$ and 0 otherwise. For self-reported flu symptoms in some cases we only have information about the interval from $t_0$ to $t_1$ in which symptoms occurred, so we assume it increases uniformly in the interval, i.e. $P_{it} = (t - t_0) / (t_1 - t_0)$.

To fit this equation we conducted a nonlinear least squares estimation procedure that utilizes the Gauss-Newton algorithm.[5] To estimate standard errors and 95% confidence intervals, we used a bootstrapping procedure in which we repeatedly re-sampled subject observations with replacement and re-estimated the fit.[6] This procedure produced somewhat wider confidence intervals than those derived from asymptotic approximations, so we report only the more conservative bootstrapped estimates of the standard errors in the Tables S3-S12.

In the left panel of Figure 2 we calculated the nonparametric maximum likelihood estimate (NPMLE) of cumulative flu incidence for both the friend group and the random group[7] and in the right panel we show the predicted daily incidence based on Model 1 in Table S3. Daily incidence for the random group is the derivative of the cumulative logistic function:

$$p_t = \lambda e^{\frac{-(t+\alpha)}{\sigma}} \bigg/ \sigma\left(1 + e^{\frac{-(t+\alpha)}{\sigma}}\right)$$

and for the friends group is:

$$p_t = \lambda e^{\frac{-(t+\alpha+\beta_{friend})}{\sigma}} \bigg/ \sigma\left(1 + e^{\frac{-(t+\alpha+\beta_{friend})}{\sigma}}\right)$$

In Figure 4, we calculate early detection days for in-degree by multiplying the coefficient and confidence intervals in Table S7 by the difference in in-degree between the above-average-in-degree group and the below-average-in-degree group. Similarly, we calculate early detection days for betweenness by multiplying the coefficient and confidence intervals in Table S9 by the difference in betweenness between the above-average-betweenness group and the below-average-betweenness group. And we calculate early detection days for transitivity by multiplying the coefficient and confidence intervals in Table S11 by the difference in transitivity between the above-average-transitivity group and the below-average-transitivity group.

**Table S1: Summary Statistics for Friend Group and Random Group**

|  | Friend Group | | Random Group | | Mann-Whitney U | |
| --- | --- | --- | --- | --- | --- | --- |
|  | Mean | S.D. | Mean | S.D. | p | N |
| Flu Diagnosis by Medical Staff | 0.075 | 0.264 | 0.078 | 0.269 | 0.876 | 744 |
| Self-Reported Flu Symptoms | 0.325 | 0.469 | 0.310 | 0.463 | 0.678 | 744 |
| In Degree | 1.435 | 0.663 | 0.433 | 0.664 | 0.000 | 744 |
| Out Degree | 2.689 | 0.543 | 2.611 | 0.672 | 0.306 | 744 |
| Betweenness Centrality (Percentile) | 0.559 | 0.271 | 0.423 | 0.294 | 0.000 | 744 |
| Transitivity | 0.142 | 0.231 | 0.148 | 0.274 | 0.039 | 721 |
| Popularity Index | 4.053 | 0.982 | 3.967 | 1.022 | 0.195 | 744 |
| Self-Reported H1N1 Vaccine | 0.200 | 0.400 | 0.188 | 0.391 | 0.685 | 744 |
| H1N1 Vaccine at UHS | 0.115 | 0.320 | 0.110 | 0.313 | 0.812 | 744 |
| Self-Reported Seasonal Flu Vaccine | 0.499 | 0.528 | 0.473 | 0.506 | 0.595 | 744 |
| Seasonal Flu Vaccine at UHS | 0.388 | 0.488 | 0.401 | 0.491 | 0.719 | 744 |
| Female | 0.720 | 0.450 | 0.627 | 0.484 | 0.007 | 744 |
| Sophomore | 0.176 | 0.382 | 0.235 | 0.425 | 0.049 | 744 |
| Junior | 0.259 | 0.439 | 0.238 | 0.427 | 0.522 | 744 |
| Senior | 0.322 | 0.468 | 0.276 | 0.448 | 0.172 | 744 |
| Varsity Athlete | 0.092 | 0.289 | 0.113 | 0.317 | 0.345 | 744 |

Friend group $N$=425, random group $N$=325. The Mann Whitney U $p$ value indicates the probability that values for the friends and random groups were drawn from the same distribution.

## Table S2: Correlates of Getting Flu by December 31, 2009

|  | *Medical Staff Flu Diagnoses* | | *Self-Reported Flu Symptoms* | |
|---|---|---|---|---|
|  | *Correlation* | *p* | *Correlation* | *p* |
| Flu Diagnosis by Medical Staff | --- | --- | 0.40 | 0.00 |
| Self-Reported Flu Symptoms | 0.40 | 0.00 | --- | --- |
| Member of Friend Group | -0.01 | 0.88 | 0.02 | 0.68 |
| In Degree | 0.01 | 0.78 | 0.08 | 0.02 |
| Out Degree | -0.01 | 0.75 | 0.01 | 0.84 |
| Betweenness Centrality | 0.02 | 0.67 | 0.03 | 0.36 |
| Transitivity | -0.03 | 0.46 | 0.05 | 0.19 |
| Popularity Index | -0.03 | 0.46 | 0.01 | 0.86 |
| Self-Reported H1N1 Vaccine | -0.04 | 0.28 | -0.03 | 0.41 |
| H1N1 Vaccine at UHS | -0.01 | 0.85 | 0.05 | 0.19 |
| Self-Reported Seasonal Flu Vaccine | 0.01 | 0.75 | 0.04 | 0.33 |
| Seasonal Flu Vaccine at UHS | 0.05 | 0.20 | 0.05 | 0.16 |
| Female | 0.02 | 0.51 | 0.06 | 0.10 |
| Sophomore | 0.04 | 0.23 | 0.04 | 0.22 |
| Junior | -0.06 | 0.09 | 0.08 | 0.02 |
| Senior | -0.06 | 0.12 | -0.07 | 0.07 |
| Varsity Athlete | 0.04 | 0.30 | -0.04 | 0.31 |

*P* values indicate probability the Pearson correlation is 0. Lack of consistent correlation suggests none of the independent variables influence overall cumulative risk of flu.

**Table S3: Effect of Being in the Friend Group on Cumulative Flu Incidence, Diagnoses by Medical Staff**

|  | Model 1 | | | | Model 2 | | | |
|---|---|---|---|---|---|---|---|---|
|  | Coef. | S.E. | Lower 95% C.I. | Upper 95% C.I. | Coef. | S.E. | Lower 95% C.I. | Upper 95% C.I. |
| *Location Variables:* | | | | | | | | |
| **Friend Group** | **-14.7** | **1.6** | **-17.6** | **-11.7** | **-15.7** | **2.1** | **-19.3** | **-12.4** |
| H1N1 Vaccination |  |  |  |  | 32.0 | 7.3 | 17.6 | 43.8 |
| Seasonal Flu Vaccination |  |  |  |  | -0.8 | 2.2 | -5.6 | 1.3 |
| Female |  |  |  |  | -8.5 | 2.2 | -12.2 | -3.2 |
| Sophomore |  |  |  |  | 25.2 | 2.9 | 20.1 | 32.1 |
| Junior |  |  |  |  | 66.7 | 3.1 | 61.2 | 72.8 |
| Senior |  |  |  |  | 57.6 | 2.5 | 54.3 | 62.3 |
| Varsity Athlete |  |  |  |  | -5.7 | 3.0 | -12.0 | -1.7 |
| Constant | 66.2 | 2.7 | 61.7 | 70.6 | 43.0 | 2.8 | 38.2 | 47.2 |
| *Scale Variable:* | 22.5 | 1.9 | 19.3 | 26.7 | 21.5 | 1.1 | 19.5 | 23.3 |
| *Residual Standard Error* |  | 0.2031 |  |  |  | 0.2022 |  |  |

Nonlinear least squares estimates of parameters in a cumulative logistic function fit to the cumulative incidence of flu diagnosed by medical staff in 744 subjects, each followed for 122 days. Location variable coefficients can be interpreted as the shift that occurs in days with respect to a unit increase in the independent variable. Standard errors and confidence intervals are bootstrapped. Results show friend group gets diagnosed with the flu by medical staff about 15 days earlier than the random group, and controlling for other factors does not affect the significance of the estimate.

**Table S4: Effect of Being in the Friend Group on Cumulative Flu Incidence, Self-Reported Data**

|  | Model 3 | | | | Model 4 | | | |
|---|---|---|---|---|---|---|---|---|
|  | Coef. | S.E. | Lower 95% C.I. | Upper 95% C.I. | Coef. | S.E. | Lower 95% C.I. | Upper 95% C.I. |
| *Location Variables:* | | | | | | | | |
| Friend Group | -3.2 | 0.6 | -4.3 | -2.2 | -2.5 | 0.6 | -3.6 | -1.6 |
| H1N1 Vaccination | | | | | 12.5 | 1.2 | 10.1 | 14.8 |
| Seasonal Flu Vaccination | | | | | 2.8 | 0.5 | 1.8 | 3.9 |
| Female | | | | | -9.0 | 0.7 | -10.3 | -7.7 |
| Sophomore | | | | | -5.3 | 0.8 | -6.9 | -4.0 |
| Junior | | | | | -7.3 | 0.6 | -8.5 | -6.2 |
| Senior | | | | | 6.9 | 0.8 | 5.6 | 8.2 |
| Varsity Athlete | | | | | 6.6 | 0.8 | 5.1 | 8.2 |
| Constant | 123.9 | 0.6 | 122.9 | 125.2 | 126.2 | 1.0 | 124.1 | 128.6 |
| *Scale Variable:* | 36.8 | 0.4 | 36.1 | 37.4 | 34.9 | 0.3 | 34.4 | 35.5 |
| *Residual Standard Error* | | | 0.3481 | | | | 0.3463 | |

Nonlinear least squares estimates of parameters in a cumulative logistic function fit to the self-reported cumulative incidence of flu in 744 subjects, each followed for 122 days. Location variable coefficients can be interpreted as the shift that occurs in days with respect to a unit increase in the independent variable. Standard errors and confidence intervals are bootstrapped. Results show the friend group self-reports flu symptoms about 3 days earlier than the random group, and controlling for other factors does not affect the significance of the estimate.

**Table S5: Effect of Self-Reported Popularity on Cumulative Flu Incidence**

|  | Model 5 (Medical Staff Diagnoses) | | | | Model 6 (Self Reports) | | | |
| --- | --- | --- | --- | --- | --- | --- | --- | --- |
|  | Coef. | S.E. | Lower 95% C.I. | Upper 95% C.I. | Coef. | S.E. | Lower 95% C.I. | Upper 95% C.I. |
| *Location Variables:* | | | | | | | | |
| Self-Reported Popularity | 3.6 | 1.0 | 2.2 | 6.0 | -1.3 | 0.2 | -1.7 | -0.8 |
| Constant | 47.4 | 4.2 | 37.7 | 52.5 | 127.2 | 1.1 | 124.8 | 128.7 |
| *Scale Variable:* | 25.5 | 1.2 | 23.4 | 27.6 | 36.8 | 0.4 | 36.2 | 37.5 |
| *Residual Standard Error* | | | 0.2032 | | | | 0.3481 | |

Nonlinear least squares estimates of parameters in a cumulative logistic function fit to the cumulative incidence of flu diagnosed by medical staff (left model) and self-reported (right model) in 744 subjects, each followed for 122 days. Location variable coefficients can be interpreted as the shift that occurs in days with respect to a unit increase in the independent variable. Standard errors and confidence intervals are bootstrapped. Results self-reported popularity has an inconsistent effect on timing of the flu.

**Table S6: Effect of Being in the Friend Group on Cumulative Flu Incidence, Controlling for Self-Reported Popularity**

|  | Model 7 (Medical Staff Diagnoses) | | | | Model 8 (Self Reports) | | | |
| --- | --- | --- | --- | --- | --- | --- | --- | --- |
|  | Coef. | S.E. | Lower 95% C.I. | Upper 95% C.I. | Coef. | S.E. | Lower 95% C.I. | Upper 95% C.I. |
| *Location Variables*: | | | | | | | | |
| *Friend Group* | -14.5 | 2.4 | -19.4 | -11.0 | -3.1 | 0.5 | -4.1 | -2.2 |
| *Self-Reported Popularity* | 4.1 | 1.0 | 2.1 | 5.5 | -1.3 | 0.3 | -1.8 | -0.7 |
| *Constant* | 53.5 | 4.2 | 48.1 | 61.8 | 128.9 | 1.2 | 127.2 | 131.4 |
| *Scale Variable:* | 24.7 | 1.0 | 23.0 | 26.3 | 36.8 | 0.4 | 36.2 | 37.6 |
| *Residual Standard Error* | 0.2031 | | | | 0.3480 | | | |

Nonlinear least squares estimates of parameters in a cumulative logistic function fit to the cumulative incidence of flu diagnosed by medical staff (left model) and self-reported (right model) in 744 subjects, each followed for 122 days. Location variable coefficients can be interpreted as the shift that occurs in days with respect to a unit increase in the independent variable. Standard errors and confidence intervals are bootstrapped. Results show the friend group gets the flu significantly earlier, even when controlling for a self-reported measure of popularity.

**Table S7: Effect of Network In Degree on Cumulative Flu Incidence**

|  | Model 9 (Medical Staff Diagnoses) | | | | Model 10 (Self Reports) | | | |
| --- | --- | --- | --- | --- | --- | --- | --- | --- |
|  | Coef. | S.E. | Lower 95% C.I. | Upper 95% C.I. | Coef. | S.E. | Lower 95% C.I. | Upper 95% C.I. |
| *Location Variables:* | | | | | | | | |
| In Degree | -5.6 | 1.3 | -8.1 | -3.6 | -8.0 | 0.4 | -8.5 | -7.3 |
| Constant | 67.8 | 1.4 | 65.4 | 70.3 | 130.2 | 0.7 | 128.6 | 131.3 |
| *Scale Variable:* | 25.2 | 1.3 | 22.8 | 27.8 | 36.8 | 0.5 | 35.9 | 37.6 |
| *Residual Standard Error* | | | 0.2032 | | | | 0.3470 | |

Nonlinear least squares estimates of parameters in a cumulative logistic function fit to the cumulative incidence of flu diagnosed by medical staff (left model) and self-reported (right model) in 744 subjects, each followed for 122 days. Location variable coefficients can be interpreted as the shift that occurs in days with respect to a unit increase in the independent variable. Standard errors and confidence intervals are bootstrapped. Results show that individuals with high in-degree tend to get the flu earlier than others.

**Table S8: Effect of Network Out Degree on Cumulative Flu Incidence**

|  | Model 11 (Medical Staff Diagnoses) | | | | Model 12 (Self Reports) | | | |
| --- | --- | --- | --- | --- | --- | --- | --- | --- |
|  | Coef. | S.E. | Lower 95% C.I. | Upper 95% C.I. | Coef. | S.E. | Lower 95% C.I. | Upper 95% C.I. |
| *Location Variables*: | | | | | | | | |
| Out Degree | 7.5 | 1.8 | 4.7 | 11.2 | 2.5 | 0.4 | 1.5 | 3.2 |
| Constant | 42.2 | 4.8 | 32.9 | 49.5 | 115.2 | 1.3 | 113.3 | 118.3 |
| *Scale Variable:* | 25.5 | 1.2 | 23.0 | 27.4 | 36.8 | 0.4 | 36.0 | 37.4 |
| *Residual Standard Error* | | 0.2032 | | | | 0.3481 | | |

Nonlinear least squares estimates of parameters in a cumulative logistic function fit to the cumulative incidence of flu diagnosed by medical staff (left model) and self-reported (right model) in 744 subjects, each followed for 122 days. Location variable coefficients can be interpreted as the shift that occurs in days with respect to a unit increase in the independent variable. Standard errors and confidence intervals are bootstrapped. Results show that the number of friends a person nominates actually *delays* the average onset of flu.

**Table S9: Effect of Betweenness Centrality on Cumulative Flu Incidence**

|  | Model 13 (Medical Staff Diagnoses) | | | | Model 14 (Self Reports) | | | |
| --- | --- | --- | --- | --- | --- | --- | --- | --- |
|  | *Coef.* | *S.E.* | *Lower 95% C.I.* | *Upper 95% C.I.* | *Coef.* | *S.E.* | *Lower 95% C.I.* | *Upper 95% C.I.* |
| *Location Variables*: | | | | | | | | |
| Betweenness Centrality | -16.5 | 8.3 | -28.3 | -1.9 | -22.9 | 1.9 | -27.2 | -20.0 |
| Constant | 62.9 | 1.1 | 60.9 | 65.0 | 123.2 | 0.5 | 122.1 | 123.9 |
| *Scale Variable:* | 25.5 | 1.1 | 23.3 | 27.5 | 36.8 | 0.4 | 35.9 | 37.4 |
| *Residual Standard Error* | | | 0.2032 | | | | 0.3479 | |

Nonlinear least squares estimates of parameters in a cumulative logistic function fit to the cumulative incidence of flu diagnosed by medical staff (left model) and self-reported (right model) in 744 subjects, each followed for 122 days. Location variable coefficients can be interpreted as the shift that occurs in days with respect to a unit increase in the independent variable. Standard errors and confidence intervals are bootstrapped. Results show that individuals with high betweenness centrality tend to get the flu earlier than others.

**Table S10: Effect of Betweenness Centrality on Cumulative Flu Incidence With Controls**

|  | Model 15 | | | | Model 16 | | | |
|---|---|---|---|---|---|---|---|---|
|  | (Medical Staff Diagnoses) | | | | (Self Reports) | | | |
|  | Coef. | S.E. | Lower 95% C.I. | Upper 95% C.I. | Coef. | S.E. | Lower 95% C.I. | Upper 95% C.I. |
| *Location Variables*: | | | | | | | | |
| *Betweenness Centrality* | -15.0 | 8.4 | -27.3 | -0.4 | -16.6 | 1.8 | -19.5 | -13.5 |
| *In Degree* | -4.2 | 1.3 | -6.2 | -1.4 | -7.6 | 0.4 | -8.3 | -6.9 |
| *Out Degree* | 7.8 | 1.7 | 5.4 | 11.4 | 3.4 | 0.5 | 2.6 | 4.3 |
| *Constant* | 46.7 | 4.7 | 37.0 | 54.0 | 121.6 | 1.3 | 118.9 | 124.3 |
| *Scale Variable:* | 25.3 | 1.2 | 23.2 | 27.5 | 36.7 | 0.4 | 35.9 | 37.4 |
| *Residual Standard Error* | 0.2031 | | | | 0.3468 | | | |

Nonlinear least squares estimates of parameters in a cumulative logistic function fit to the cumulative incidence of flu diagnosed by medical staff (left model) and self-reported (right model) in 744 subjects, each followed for 122 days. Location variable coefficients can be interpreted as the shift that occurs in days with respect to a unit increase in the independent variable. Standard errors and confidence intervals are bootstrapped. Results show that betweenness centrality remains a significant predictor of early flu onset even when controlling for degree variables.

**Table S11: Effect of Transitivity on Cumulative Flu Incidence**

|  | Model 17 (Medical Staff Diagnoses) | | | | Model 18 (Self Reports) | | | |
| --- | --- | --- | --- | --- | --- | --- | --- | --- |
|  | Coef. | S.E. | Lower 95% C.I. | Upper 95% C.I. | Coef. | S.E. | Lower 95% C.I. | Upper 95% C.I. |
| *Location Variables*: | | | | | | | | |
| Transitivity | 31.9 | 4.8 | 23.5 | 43.5 | 15.0 | 1.6 | 12.7 | 18.5 |
| Constant | 56.9 | 1.5 | 53.5 | 59.0 | 153.9 | 1.2 | 151.3 | 155.8 |
| *Scale Variable:* | 24.8 | 0.8 | 23.3 | 26.6 | 40.5 | 0.7 | 39.1 | 41.7 |
| *Residual Standard Error* | | | 0.2046 | | | | 0.2873 | |

Nonlinear least squares estimates of parameters in a cumulative logistic function fit to the cumulative incidence of flu diagnosed by medical staff (left model) and self-reported (right model) in 721 subjects, each followed for 122 days. Location variable coefficients can be interpreted as the shift that occurs in days with respect to a unit increase in the independent variable. Standard errors and confidence intervals are bootstrapped. Results show that individuals with low transitivity tend to get the flu earlier than others.

**Table S12: Effect of Transitivity on Cumulative Flu Incidence with Controls**

|  | Model 19 (Medical Staff Diagnoses) | | | | Model 20 (Self Reports) | | | |
| --- | --- | --- | --- | --- | --- | --- | --- | --- |
|  | Coef. | S.E. | Lower 95% C.I. | Upper 95% C.I. | Coef. | S.E. | Lower 95% C.I. | Upper 95% C.I. |
| *Location Variables:* | | | | | | | | |
| Transitivity | 34.9 | 4.2 | 25.1 | 42.0 | 22.8 | 1.8 | 19.1 | 25.6 |
| In Degree | -3.6 | 1.2 | -5.4 | -1.0 | -4.6 | 0.5 | -5.6 | -3.8 |
| Out Degree | 17.6 | 2.2 | 13.7 | 21.3 | 17.3 | 0.7 | 16.1 | 19.0 |
| Constant | 13.2 | 6.0 | 2.2 | 21.1 | 109.7 | 1.9 | 106.9 | 112.8 |
| *Scale Variable:* | 25.0 | 1.0 | 23.1 | 26.9 | 39.4 | 0.7 | 38.1 | 41.0 |
| *Residual Standard Error* | | | 0.2045 | | | | 0.2860 | |

Nonlinear least squares estimates of parameters in a cumulative logistic function fit to the cumulative incidence of flu diagnosed by medical staff (left model) and self-reported (right model) in 721 subjects, each followed for 122 days. Location variable coefficients can be interpreted as the shift that occurs in days with respect to a unit increase in the independent variable. Standard errors and confidence intervals are bootstrapped. Results show that transitivity remains a significant predictor of early flu onset even when controlling for degree variables.

**REFERENCES for SUPPLEMENTARY INFORMATION**